\documentclass[extra,onecolumn]{gji}
\usepackage{graphicx,epsf,subfigure}
\usepackage{amssymb}

\title[Geophys.\ J.\ Int.: Tidal instability in a rotating and differentially heated ellipsoidal shell]
  {Tidal instability in a rotating and differentially heated ellipsoidal shell}

\author[D. C\'ebron, P. Maubert, M. Le Bars]
  {D. C\'ebron\thanks{corresponding author (cebron@irphe.univ-mrs.fr)}, P. Maubert, M. Le Bars \\
   Institut de Recherche sur les Ph\'enom\`enes Hors Equilibre, UMR 6594, \\
CNRS et Aix-Marseille Universit\'e, 49 rue F. Joliot-Curie, BP146, 13384 Marseille Cedex 13, France.}

\date{Received 2010 February 23; in original form 2010 February 23}
\volume{200}
\pubyear{2010}

\newcommand{\D}{\mathrm{d}}

\begin{document}

\label{firstpage}

\maketitle

\begin{summary}
The stability of a rotating flow in a triaxial ellipsoidal shell
with an imposed temperature difference between inner and outer
boundaries is studied numerically. We demonstrate that (i) a stable
temperature field encourages the tidal instability, (ii) the tidal
instability can grow on a convective flow, which confirms its
relevance to geo- and astrophysical contexts, and (iii) its growth
rate decreases when the intensity of convection increases. Simple
scaling laws characterizing the evolution of the heat flux based on
a competition between viscous and thermal boundary layers are
derived analytically and verified numerically. Our results confirm
that thermal and tidal effects have to be simultaneously taken into
account when studying geophysical and astrophysical flows.
\end{summary}



\section{Introduction} \label{intro}

The tidal or elliptical instability comes from a triadic parametric
resonance between two inertial waves of a rotating fluid and an
imposed elliptic deformation \cite[see for instance the review
by][and references therein]{kerswell_2002}. It is a generic
phenomenon in turbulence \cite[e.g.][]{Widnall_1974,Moore_1975} and
vortex dynamics \cite[e.g.][]{Leweke98_b,LeDizes_2002,Meunier_2002}.
There, the elliptical deformation comes from interactions between
adjacent vortices. But this instability is also studied in
geophysical and astrophysical tidally deformed bodies: for example,
its presence has been suggested (i) in binary stars
\cite[][]{Rieutord,LeBars09} and accretion disks
\cite[][]{Goodman_1993,Ryu_1994}, where they could participate in
the energy and angular momentum exchanges between neighbouring
systems \cite[][]{Lubow_1993}, and (ii) in planetary cores
\cite[e.g.][]{Aldridge, Cebron_2009}, where they could play an
important role in the induction of a magnetic field
\cite[][]{KerswellMalkus, Lacaze, Herreman_2009} and even in
planetary dynamos \cite[][]{Malkus93}.

Previous studies of tidal instability have been performed for an
isothermal fluid, starting from a solid body rotation or from an
homogeneous isotropic turbulent rotating flow
\cite[][]{Fabijonas_2003}. Nevertheless, in all natural systems,
temperature differences are also present, which lead either to
stratification or to convection. The previous studies need then to
be reinvestigated in order to quantify the impact of a temperature
field on the elliptical instability. Conversely, from a thermal
point of view, many studies have been performed regarding the
convective flow of an incompressible homogeneous fluid in a rotating
spherical shell \cite[see for instance][and references
therein]{Aurnou_2007}. However, since most astrophysical bodies are
tidally deformed, these studies have to be reinvestigated to account
for the potential presence of an elliptical instability. Note that a
link between tidal resonance of gravito-inertial waves and thermal
effects has
 been proposed by \cite{Kumazawa_1994} in a geophysical context.
However, this study considers the resonant forcing of a given wave
by tides, which is different from the elliptical instability
possibly induced by parametric resonance. Then, to the best of our
knowledge, the only previous work which combines thermal and
elliptical effects is the theoretical study of \cite{LeBars06}, who
considered the influence of an established diffusive temperature
profile. They then interpreted the thermo-elliptical instability in
terms of gravito-inertial waves resonance and determined its linear
growth rate by a local approach. In order to complete this work, a
guideline of our study is given by the following questions: (i) how
is the growth rate of the tidal instability modified by an imposed
temperature difference in the spheroidal geometry? (ii) Does the
tidal instability grow over an established convective flow? (iii) Is
the heat flux modified by the instability, and what are the scaling
laws involved? In this paper, we tackle these questions using a
systematic numerical study of the thermo-elliptical rotating flow in
a fixed triaxial ellipsoidal shell. The numerical model is presented
in section 2. The variations of the heat flux and the growth rate
are systematically quantified and interpreted in section 3. Finally,
orders of magnitude and conclusions are drawn for geo- and
astrophysical flows in section 4.

\section{Numerical model and its validation} \label{section1}

\begin{figure}
  \begin{center}
    \epsfysize=11.0cm
    \leavevmode
    \epsfbox{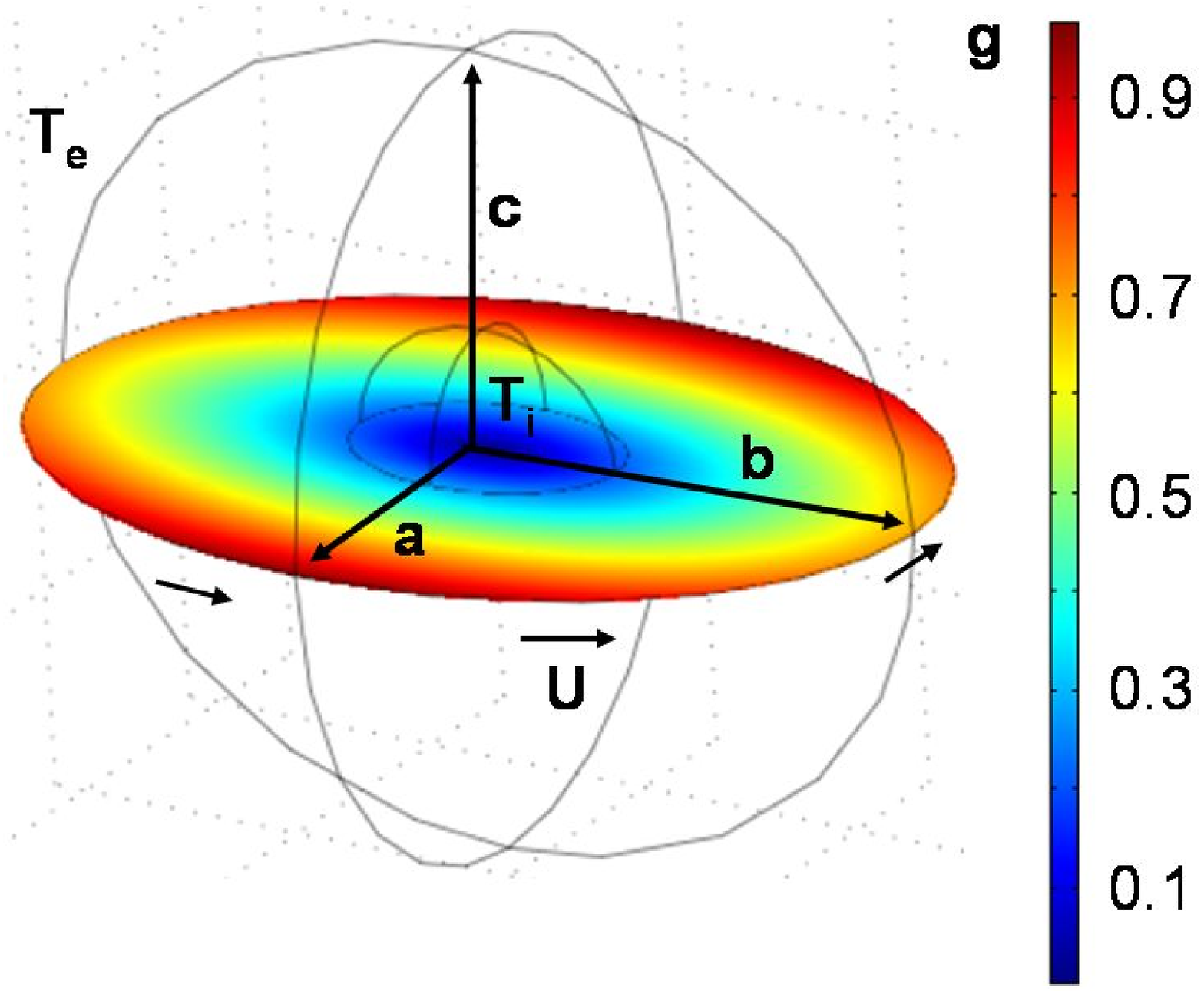}
    \caption{\it{}}
    \label{cebronfig1}
  \end{center}
\end{figure}

As sketched in figure \ref{cebronfig1}, we consider the rotating
flow inside a fixed triaxial ellipsoid of axes $(a,b,c)$, with a
constant tangential velocity $U\sqrt{1-(z/c)^2}$ imposed all along
the outer boundary in each plane of coordinate $z$ perpendicular to
the rotation axis $(Oz)$, where $U$ is the imposed boundary velocity
at the equator. An homothetic ellipsoidal core, in a ratio
$\eta=0.3$, is placed at the center, with also an imposed constant
tangential velocity $\eta U\sqrt{1-(z/c)^2}$ along its boundary. In
the following, we use the mean equatorial radius $R_{eq}=(a+b)/2$ as
a lengthscale and we introduce the timescale $\Omega^{-1}$ by
writing the tangential velocity along the deformed outer boundary at
the equator $U=\Omega R_{eq}$. Then, the problem is fully described
by the adimensionalized Navier-Stokes equations with four
dimensionless numbers: the eccentricity
$\varepsilon=(b^2-a^2)/(a^2+b^2)$, the inner to outer core ratio
$\eta$, the aspect ratio $c/b$, and the Reynolds number $Re=\Omega\
R_{eq}^2/\nu$, or its inverse the Ekman number $E_k$, where $\nu$ is
the kinematic viscosity of the fluid. In all the computations
presented here, we choose $c$ equal to the mean equatorial radius
$R_{eq}=(a+b)/2$, as in classical experimental setups (see e.g.
\cite{lacaze_2005}). This allows us to focus on the well-known
stationary spin-over mode of the tidal instability. Note that for
lots of tidally deformed planetary and stellar bodies, the c-axis is
the shortest axis because of the rotational flattening, whereas the
longest axis is the one pointing towards the tide-raising body. In
this configuration, the first selected mode of the tidal instability
close to threshold may change (\cite{KerswellMalkus},
\cite{Cebron_2009}), giving rise to oscillatory motions.
Nevertheless, we expect the main physical processes described here
to remain valid for these other modes.

In addition to these purely hydrodynamic considerations, we also
solve the temperature equation with constant and uniform
temperatures $T_i$ and $T_e$ applied respectively at the inner and
outer boundaries. To model the buoyancy, the density is assumed to
vary linearly with the temperature and the Boussinesq approximation
is used. Then, the gravity term of the Navier-Stokes equation is
written: $\rho\ \mathbf{g}_*=\rho(T_e)\ [1-\alpha (T_*-T_e)]\
\mathbf{g}_*$, where $\rho$ is the fluid density, $\alpha$ the
coefficient of thermal volumic expansion, $\mathbf{g}_*$ the local
gravity which depends on the position and $T_*$ the local
temperature. In our model, the gravity is calculated by solving the
Poisson equation for the gravitational potential $\phi$, assuming
that the outer boundary is an isopotential. Thus, the thermal part
of the problem is controlled by two dimensionless numbers: the
Prandtl number $Pr=\nu / \kappa$, where $\kappa$ is the thermal
diffusivity, and the Rayleigh number $Ra=(\alpha\ g_0\ (T_i -
T_e)\ d^3)/(\kappa\ \nu)$, where $d=R_{eq} (1-\eta)$ is the shell
thickness and $g_0$ is the surface gravity at the outer boundary of
the corresponding sphere, i.e. $g_0=4/3\ \pi\ G\ R_{eq}\ \rho(T_e)$ with
$G$ the gravitational constant. Note that the kinematic viscosity of
the fluid is assumed independent of the temperature in this work.

Finally, the problem solved is described by the following system of
dimensionless equations:
\begin{eqnarray}
\frac{\partial \mathbf{u}}{\partial t}+ \mathbf{u} \cdot
\mathbf{\nabla} \mathbf{u} = -\mathbf{\nabla} p + \frac{1}{Re}\
\mathbf{\bigtriangleup} \mathbf{u}  - \frac{Ra}{Pr\ Re^2\ (1-\eta)^3}
T\ \mathbf{g},
\end{eqnarray}
\begin{eqnarray}
\mathbf{\nabla}  \cdot \mathbf{u} =0,
\end{eqnarray}
\begin{eqnarray}
\frac{\partial T}{\partial t}+ \mathbf{u} \cdot \mathbf{\nabla} T =
\frac{1}{Pr\ Re}\ \mathbf{\bigtriangleup} T,
\end{eqnarray}
\begin{eqnarray}
\bigtriangleup \phi = 3,
\end{eqnarray}
\begin{eqnarray}
\mathbf{g} = - \mathbf{\nabla} \phi,
\end{eqnarray}
where $T$ is the dimensionless temperature anomaly
$T=(T_*-T_e)/(T_i-T_e)$. All boundary conditions have been described
above. This numerical setup is a simple model for a tidally deformed
liquid planetary core with a solid inner core and an imposed
temperature contrast. Note that in a geophysical context, because of
the compressibility, the temperature difference $\bigtriangleup
T=T_i - T_e$ to consider is the superadiabatic temperature
difference between the inner core boundary and the core-mantle
boundary (CMB). All dimensionless numbers as well as explored
numerical ranges and typical values for the Earth's and Io's cores
are given in table \ref{tableNSD}.

Since there is no simple symmetry in the studied configuration, the
fast and precise spectral methods classically used in planetary
cores studies are not relevant here. Our computations are performed
with a commercial software using a finite element method, which
allows to correctly reproduce the geometry and to simply impose the
boundary conditions. An unstructured mesh with tetrahedral elements
was created. The mesh element type employed is the standard Lagrange
element $P1-P2$, which is linear for the pressure field but
quadratic for the velocity field. The temporal solver is IDA
\cite[][]{Hindmarsh_2005}, based on backward differencing formulas.
At each time step the system is solved with the sparse direct linear
solver PARDISO\footnote{www.pardiso-project.org}.

The numerical procedure is the same for all the computations
presented in this paper: starting from a fluid at rest in a
spherical shell of radius $R_{eq}$ with a diffusive temperature
profile, the shell is set in rotation at a constant angular velocity
$\Omega$ and a first statistically stationary state is obtained by
solving the Navier-Stokes and temperature equations; then, the
spherical shell is instantaneously deformed to the chosen triaxial
ellipsoidal shape and we follow the dynamics. Note that this
instantaneous deformation allows to study the growth rate of the
elliptical instability on a well-known conductive or convective base
flow. But we have also performed computations with a gradual
increase of the deformation, as well as computations starting from a
deformed shell with a fluid at rest before imposing the rotation at
the boundaries: the classical isothermal tidal instability then
grows, which is only slightly modified by then imposing a
temperature difference. In all cases, the final steady-state is the
same.

Two quantities are systematically studied: (i) the growth rate
$\sigma$ of the tidal instability, which quantifies the feedback of
the temperature field on the elliptical instability, and (ii) the
Nusselt number $Nu$, which quantifies the feedback of the elliptical
instability on the thermal properties of the system. Since the
elliptical instability induces a strong three-dimensional
destabilization of the initial flow, the growth rate $\sigma$ is
determined as the time constant of the exponential growth of the
mean value of the vertical velocity $W=\frac{1}{V} \int_{V} |w|\ \D
\tau $, where $w$ stands for the dimensionless vertical velocity and
$V$ for the volume of the ellipsoidal shell. The Nusselt number is
calculated by dividing the measured mean heat flux at the outer
boundary (once a statistically stationary state is reached) by the
theoretical value of the conductive heat flux in a sphere of radius
$R_{eq}$. It has systematically been checked that the difference
between this analytically known flux and the numerically calculated
flux for a purely conductive ellipsoid is below $2 \%$ for our all
simulations.

The purely hydrodynamic part of the numerical simulation has already
been validated in \cite{Cebron_2009}. A complementary validation of
the resolution of thermal effects has been done by computing the
heat flux in a rotating spherical shell: in figure \ref{cebronfig2}
(a), our results at $Re=344$ are compared with the compilation of
results given in \cite{Aurnou_2007} for the modified Nusselt number
defined by $Nu^*=\frac{Nu\ E}{Pr}$, where $E=\frac{\nu}{\Omega\ d^2}$
is the Ekman number based on the gap width, i.e. $E=((1-\eta)^2\
Re)^{-1}$. Note that our numerical results fill an area with few
data and agree with the general trend. Besides, our results are
compared in figure \ref{cebronfig2} (b) with the scaling law
proposed by \cite{Christensen_2002} obtained from numerical simulations:
\begin{eqnarray}
Nu^* = 0.077\ {(Ra^*_Q)}^{5/9} \label{eq:Nuetoile}
\end{eqnarray}
where $Ra^*_Q=Ra\ Nu\ E^3/Pr^2$ is the Rayleigh number based on the
heat flux instead of the temperature contrast and which does not
take into account any viscous or thermal diffusive effect. An
excellent quantitative agreement is found.

\section{Interaction between the elliptical instability and the thermal field}\label{interaction}

In their analytical WKB analysis of the linear stability of
elliptical streamlines in the presence of a diffusive temperature
field, \cite{LeBars06} found an analytical expression of the
inviscid growth rate of the elliptical instability in the limit of
small deformations as:
\begin{eqnarray}
\sigma=\frac{9-3 \tilde{Ra}}{16-4 \tilde{Ra}}\ \varepsilon ,\label{eq:WKB}
\end{eqnarray}
where $\tilde{Ra}=\frac{\eta-1}{\ln(\eta)} \frac{Ra\ E^2}{Pr}$ is a
modified Rayleigh number. The expression \ref{eq:WKB} reflects the
modification by thermal effects of the elliptical instability, which
in this context corresponds to the parametric resonance of 2
gravito-inertial waves with the tidal deformation. \cite{LeBars06}
also showed that the elliptical instability cannot grow for
$\tilde{Ra}>3$ because no resonance between inertial modes is then
possible. As shown in figure \ref{cebronfig4} (a), the same
decreasing trend with $Ra$ is recovered in the spherical shell, even
if the results differ quantitatively from this theoretical local
estimate. The first noticeable result from our complete calculations
is that the growth rate of the elliptical instability is
significantly enhanced by the thermal stratification (i.e. $Ra <
0$), as first suggested by \cite{LeBars06}. It may seem at first
surprising that an a priori stabilizing effect leads to an
encouragment of flow destabilization, but such a behavior has
already been shown for instance in a Taylor-Couette flow, where a
stable configuration can be destabilized by the presence of an axial
stratification, leading to the so-called strato-rotational
instability \cite[see for instance the recent experiments
by][]{LeBars07}. The second result of planetary relevance is that
the tidal instability can still grow on an established convective
flow. As shown in figure \ref{BusseIE}, the organization of the flow
is then completely changed. Indeed, in the absence of tidal
instability, the convective flow organizes in a classical columnar
structure shown in figure \ref{BusseIE} (a) (the so-called Busse
columns), where motions are mainly bidimensional and vertical
velocities remain small, coming only from the Ekman pumping inside
the columns. On the contrary in figure \ref{BusseIE} (b), Busse
columns are completely destroyed by the tidal instability, whose
velocity field is fully three-dimensional. The growth rate $\sigma$
of the elliptical instability progressively decreases when the
Rayleigh number increases as shown in figure \ref{cebronfig4} (a).
Indeed, the thermal convective motions can be seen as a perturbation
superimposed on the classical isothermal base flow of the elliptical
instability, corresponding to a rotation along elliptical
streamlines: when the Rayleigh number increases, the typical
convective velocity increases and the ellipticity of the base flow
streamlines is less and less felt by the fluid particles. The
dimensionless version of this model can be quantified in defining
the typical Rossby number of the flow, i.e. the ratio between the
root mean square convective velocity and the typical rotating
velocity $\Omega R_{eq}$ : $Ro=\sqrt{\frac{1}{V} \int \mathbf{u}
\cdot \mathbf{u}\ \D \tau}$, where $\mathbf{u}$ is the dimensionless
velocity in the rotating frame. One would expect to find a critical
Rossby number of order 1 for the disappearance of the tidal
instability. As shown in figure \ref{cebronfig4} (b), the critical
value $Ro \approx 0.25$, where the growth rate is zero, is indeed
found for $Pr=1$ and $Re=344$, which is near the threshold of the
tidal instability. The critical Rossby number then seems to increase
with the Reynolds number (see figure \ref{cebronfig4} (a) for
$Re=688$), if it exists at all. Indeed, when the Rayleigh number
increases, the typical convective velocity also increases whereas
the typical lengthscale of convection $L_{conv}$ simultaneously
decreases. In the limit of very large Rayleigh numbers, $L_{conv}$
is very small and turbulent motions become fully tridimensional. We
then expect thermal convective motions to be seen as an effective
turbulent viscosity, where the results of \cite{Fabijonas_2003}
predict an encouragement of the tidal destabilization. The extensive
study of this regime is very interesting, but currently out of reach
of our numerical resources. Nevertheless, our current results allow
to use the critical Rossby number $Ro_c=0.25$ determined previously
for $Re=344$ as a lower bound for flow at larger Reynolds numbers.
This result is sufficient for the planetary applications that we are
interested in, as will be shown in the next section.

When thermal and tidal instabilities are simultaneously present,
heat transfers are determined by the competition between the natural
thermal convection and the forced convection due to the elliptical
instability. In the latter case, the flow driven by the instability
can be schematically described by three distinct regions: a fully
mixed and isothermal bulk, and viscous boundary layers at the outer
and inner boundaries, where the temperature is purely diffusive. The
advective heat flux driven by the tidal instability is then
determined by the size of the viscous boundary layers
$\delta_{\nu}$, following $Nu-1\sim d/ \delta_{\nu}$. With the
classical scaling law of viscous boundary layers, we obtain $Nu-1 =
\alpha_1/ \sqrt{E^*}$, where $\alpha_1$ is a constant,
$E^*=\frac{\nu}{u_{ei}\ d}$ the Ekman number associated with the flow
driven by the instability, and $u_{ei}$ the fluid velocity due to
the elliptical instability. As shown in \cite{Cebron_2009}, this
velocity $u_{ei}$ may be computed numerically in the absence of
temperature gradient by subtracting to the computed flow the
theoretical base flow corresponding to an imposed rotation along
elliptical streamlines. $\frac{u_{ei}}{\Omega\ R_{eq}} $ generally scales
as $\alpha_2 \sqrt{Re-Re_c} $ near the threshold of instability,
where $\alpha_2 \approx 0.036$ has been determined numerically and
where $Re_c$ is the critical Reynolds number for the onset of tidal
instability equal to $Re_c=(5.24/\varepsilon)^2$ in the isothermal case for a 
stationary deformation.
$\frac{u_{ei}}{\Omega\ R_{eq}}$ then saturates toward $1$ far from the
threshold. Keeping the same scaling in the presence of thermal
effects, we expect the Nusselt number to vary as:
\begin{eqnarray}
Nu-1=\alpha_1 \sqrt{\alpha_2\ (1-\eta)}\ \left(\frac{Re}{Re_c}-1 \right)^{1/4} \sqrt{Re}.
\label{scalingflux}
 \end{eqnarray}
Our numerical results shown in figure \ref{cebronfig5} (a) validate
this scaling with $\alpha_1 \approx 0.01$. Far from threshold,
$E^*=E$ and the scaling law becomes
\begin{eqnarray}
Nu = \alpha_1/ \sqrt{E}. \label{heatlaw}
\end{eqnarray}
Note that according to this model, the Nusselt number is a measure
of the amplitude of the instability. Then, the agreement between our
results and the scaling law confirms that the amplitude of the
instability is not inhibited by a strong stratification, provided
that the Reynolds number is far enough from the threshold.

The systematic evolution of the Nusselt number with the Rayleigh
number is shown in figure \ref{cebronfig5} (b). In the absence of
tidal and convective instabilities, $Nu$ remains equal to 1. In the
presence of tidal instability, the heat flux remains constant at a
value larger than 1 up to a transition Rayleigh number which depends
on $Re$, where the natural convection becomes more efficient than
the forced one. Indeed, (\ref{scalingflux}) and (\ref{heatlaw})
describe the heat flux by the forced convection only, and are valid
in particular for stably-stratified regimes and before the onset of
convection $Ra<Ra_c$. But following \cite{King_2009}, once $Ra>Ra_c$
and convective and tidal instabilities are competing, the heat
transfer is determined by the most efficient mechanism. Following
scaling laws (\ref{eq:Nuetoile}) and (\ref{heatlaw}), the transition
between the forced convection and the natural convection appears for
the transition Nusselt number $Nu^*_t= \frac{Nu\ E}{Pr} = 0.077\
{(Ra^*_Q)}^{5/9} = \alpha_1\ \frac{\sqrt{E}}{Pr}$, which corresponds
to the transition Rayleigh number:
\begin{eqnarray}
Ra_t \approx 2.5\ E^{-8/5}\ Pr^{1/5}.
\end{eqnarray}
For $Ra > Ra_t$, heat transfers are controlled by the natural
convective motions, whereas for $Ra < Ra_t$, heat transfers are
controlled by the tidal instability. For planetary applications, it
is more convenient to write this result in term of flux Rayleigh
number defined by ${Ra}_f=Ra\ Nu$, for which the transition appears
at
\begin{eqnarray}
{Ra_f}_t \approx 0.025 \ E^{-21/10}\ Pr^{1/5}.  \label{eq:Raft}
\end{eqnarray}

\section{Planetary and stellar applications}\label{quantif}

As explained in details by \cite{Sumita_2002}, a stable density
stratification may have existed in the whole early Earth outer core
and could even have been sufficiently large to prevent dynamo
action. This initial stratification has been disrupted by some
mechanism, which could have been either a gradual mechanism,
including encroachment and entrainment \cite[][]{Lister_1998}, or
catastrophic mechanisms driven for instance by the 1/3-annual forced
nutation \cite[][]{Williams_1994} or by tidal resonant forcing of
inertial-gravity waves \cite[][]{Kumazawa_1994}. Our work suggests
an alternative proposal based on a parametric triadic resonance of
gravito-inertial waves. Note that since the stratification enhances
the growth rate of the tidal instability and since the early Earth
was already unstable from a purely hydrodynamic analysis
\cite[][]{Cebron_2009}, then a fortiori with a thermal
stratification, the tidal instability can be a mechanism for this
overturn, hence playing a role in one of the major events in
geological Earth history.

We can also apply the heat transfer scaling laws obtained in the
previous section to the present convective state of the Earth's
core. Following \cite{Christensen_2006}, the typical values are $E
\approx 5 \cdot 10^{-15}$, $Ra^*_Q \approx 3 \cdot 10^{-13}$ and
${Ra_f \approx 10^{29}} $ for $Pr=0.25$. The first consequence that
we can deduce is that $Ro \sim 10^{-6}$ according to the scaling law
 $Ro = 0.85\ {(Ra^*_Q)}^{0.41}$ given by \cite{Christensen_2006}, which
is significantly below the lower bound for disappearance of
elliptical instability $Ro_c \approx 0.25$ found in section
\ref{interaction}. That means that the convection does not prevent
the development of the tidal instability. Besides, the scaling law
(\ref{eq:Raft}) gives the estimate ${Ra_f}_t \sim 2\times 10^{28}$.
The Earth's core is just above the transition between the two kind
of convections considered in section \ref{interaction} and its heat
flux is thus controlled by natural convection. However, the
closeness of the values obtained here does not preclude a tidally
dominated flux in previous times or in other planetary systems, and
in any case highlights the relevance of the elliptical instability
as one of the major process at work in planetary cores.

To finish with, many authors have assumed that the top of the Earth
core is stably stratified, this layer being the so-called ocean of
the core (see for instance \cite{Stanley_2008} for a recent review
of the literature on this subject). Actually, as described in this
recent work, many clues lead to the idea that several planets
contain stably stratified layers in their electrically conducting
regions. An opposite (but dynamically similar) situation occurs in
stars under the photosphere, where the stratified radiative zone
takes place under the so-called convective zone. Then, we can
address the simple following questions: in such a bi-layer flow, can
the elliptical instability grow? In which area, i.e. the stratified
zone, the convective zone or both? And is the heat flux completely
modified by the instability? A single simulation is presented in
figure \ref{cebronfig7} as an illustration of the answer. In this
simulation, the same temperature $T=0$ is imposed at boundaries, and
a temperature $T=1$ is imposed at an internal neutral boundary
corresponding to an homothetic ellipsoid in a ratio $\eta_2=0.7$ :
this allows to obtain a stratified inner layer and a convective
outer layer, with clearly visible convective cells (figure
\ref{cebronfig7} (a)). Once the tidal instability is excited, figure
\ref{cebronfig7} (b) shows that the instability grows over the whole
fluid and controls the heat flux. Hence in presence of the
elliptical instability, a thermal stratification (i.e. a
subadiabatic temperature profile) is not synonymous of thermal
isolation, which is a key point from a geophysical point of view.

\section{Conclusion}

This paper presents the first numerical study of the interaction
between the tidal instability and a thermal field. Even if the range
of parameters (e.g. the Reynolds number) accessible to our numerical
tool is relatively limited, we have validated general physical
processes of direct relevance for planetary dynamics. We have shown that
a stable temperature field encourages the tidal instability. We have
demonstrated also that the tidal instability
can grow on a convective flow, and may disrupt the famous Busse
columns in planetary cores. Finally, we have demonstrated that the heat flux at
planetary scales may be controlled by the forced convection due to this tidal
instability, which in any case plays a fundamental role in the organization of fluid motions.

\newpage

\begin{table}
 \caption{List of relevant dimensionless parameters with the explored numerical ranges as well as their typical values in the Earth's and Io's cores.}\label{tableNSD}

 \label{symbols}
 \begin{tabular}{@{}lcccccc}
    Dimensionless parameters & Definition & Numerical ranges
        & Earth \& Io's core estimates  \\
  Eccentricity & $\varepsilon=\frac{b^2-a^2}{a^2+b^2}$ & $0\ -\   0.32$
        & $10^{-7}\ -\   10^{-4}$ \\
  Core radius ratio & $\eta$ & $0\ -\  0.3$
        & $0.35\ -\ ?$  \\
  Oblateness & $c/b$ & $0.86\ -\   1$
        & $997/100\ -\ 995/100$ & \\
  Reynolds number & $Re=\Omega\ R_{eq}^2/\nu$ & $0 \leq 10^{3}$
        & $10^{15}\ -\ 10^{13} $ \\

  Ekman number & $E_k=1/Re$ & $ \geq 0.001 $ & $10^{-15}\ -\ 10^{-13} $  \\
  Ekman number based on the gap & $E=((1-\eta)^2\ Re)^{-1}$  & $\geq 0.002 $ & $10^{-15}\ -\ 10^{-13} $  \\
 Prandtl number & $Pr=\nu / \kappa$ & $1$ & $0.25\ -\ 0.25$  \\
 Rayleigh number & $Ra=\frac{\alpha\ g_0 \ (T_i - T_e)\ d^3}{\kappa\ \nu}$ & $-3.8 \cdot 10^{5}\ -\ 10^{6}$ & $10^{23}\ -\ ?\ (< Ra_c)$  \\
 Flux-based Rayleigh number & $Ra^*_Q=Ra\ Nu\ E^3/Pr^2$ & $-2.6 \cdot 10^{-5}\ -\   0.24$ & $10^{-13}\ -\ ?$ \\
 Nusselt number & $Nu=Q_{total}/Q_{conduction}$ & $1\ -\ 5.6$ & $10^{6}\ -\ ? $  \\
 Modified Nusselt number & $Nu^*=\frac{Nu\ E}{Pr}$ & $\geq 0.002$ & $10^{-8}\ -\ ? $  \\
 Rossby number & $Ro=\frac{\sqrt{\frac{1}{V} \int U^2\ \D \tau}}{\Omega\ R_{eq}^2}$ & $0\ -\  0.4$ & $10^{-6}\ -\ ? $ \\
 \end{tabular}

 \medskip
\end{table}

\newpage
\clearpage

\bf{Figure 1.} \it{Schematic representation of the considered
problem. The gravity field shown here in the equatorial plane is
calculated explicitly, assuming that the outer boundary is an
isopotential.}

\bigskip

\bf{Figure 2.} \it{Evolution of the modified Nusselt number in the case of a rotating spherical shell ($\varepsilon=0$) for $Pr=1$, $\eta=0.3$ and $Re=344$.
    (a) Our computations are compared with a compilation of numerical and experimental results given by \cite{Aurnou_2007}.
    (b) Our computations are compared with the scaling law (\ref{eq:Nuetoile}) given by \cite{Christensen_2002}.}

\bigskip

\bf{Figure 3.}  \it{(a) Influence of the Rayleigh number on the growth rate of the tidal instability for $\varepsilon=0.317 $, $c=\frac{a+b}{2}$,
    $Pr=1$ and $Re=344$ or $Re=688$.  The growth rate given by equation (\ref{eq:WKB}) is also shown, taking into account a classical
    damping by a surfacic viscous term $-K/ \sqrt{E_k}$, where $K$ is a constant of order $1$ \cite[][]{lacaze_2005}.
    The values of $K$ have been determined by matching the
    theoretical value with the numerical one at $Ra=0$, giving $K=3.07$ for $Re=344$ and $K=2.91$ for $Re=688$. (b) Evolution of the growth rate of the
tidal instability with the Rossby number, calculated following the
scaling law given by \cite{Christensen_2006} $Ro = 0.85\
{(Ra^*_Q)}^{0.41}$. For $Re=344$, the instability disappears for $Ro
\approx 0.25$. }

\bigskip

\bf{Figure 4.} \it{Comparison between the convective flows in the absence of tidal instability (a, $\varepsilon = 0$) and in the presence of tidal instability (b, $\varepsilon = 0.317$) for the same values of the Reynolds
    number $Re=344$, the Rayleigh number $Ra=18762$ and the Prandtl number $Pr=1$. The vertical velocity is shown in three different planes. Also shown in (b) is an iso-surface of the norm of the velocity $||\mathbf{u}||=0.19 $, which clearly exhibits the 'S' shape of the spin-over mode.}

\bigskip

\bf{Figure 5.} \it{(a) Influence of the Reynolds number on the
Nusselt number for a convective flow with $\varepsilon=0.317 $,
$c=\frac{a+b}{2}$, $Pr=1$ and $Ra=18762$.
 (b) Evolution of the Nusselt number with the Rayleigh number in
    spherical and ellipsoidal ($\varepsilon=0.317 $, $c=\frac{a+b}{2}$) shells for $Pr=1$ and $Re=344$ or
    $Re=688$. Note that the classical dependance of the critical Rayleigh for the convection $Ra_c\propto Re^{4/3}$ (see \cite{Tilgner_1997} for instance) can be observed: the decrease of the Reynods number modifies the threshold and then, increases the convective motions.}
\bigskip

\bf{Figure 6.} \it{Visualization of the reduced temperature $T$ in
the case of a stratified layer under a convective zone ($Re=344$,
$c=\frac{a+b}{2}$).
    The inner and outer boundaries verify $T=0$ and at a ratio $\eta_2=0.7$, a temperature $T=1$ is imposed such that the Rayleigh
    number based on the thickness of the shell is $Ra=18762$, whereas the Rayleigh number based on  $\eta_2$ is equal to
    $Ra_2=Ra \frac{(1-\eta_2)^3}{(1-\eta)^3} \approx 1512$. (a) Temperature in the equatorial plane for a spherical shell of outer radius
    $R_{eq}=c$. (b) Temperature in the equatorial plane and surface of iso-value $||\mathbf{u}||=0.19 $ at saturation of the tidal instability
    for the corresponding ellipsoidal shell at $\varepsilon=0.317$. }

\newpage

\clearpage

\bigskip

\newpage
\clearpage

\begin{figure}
  \begin{center}
    \epsfysize=10.0cm
    \leavevmode
    \epsfbox{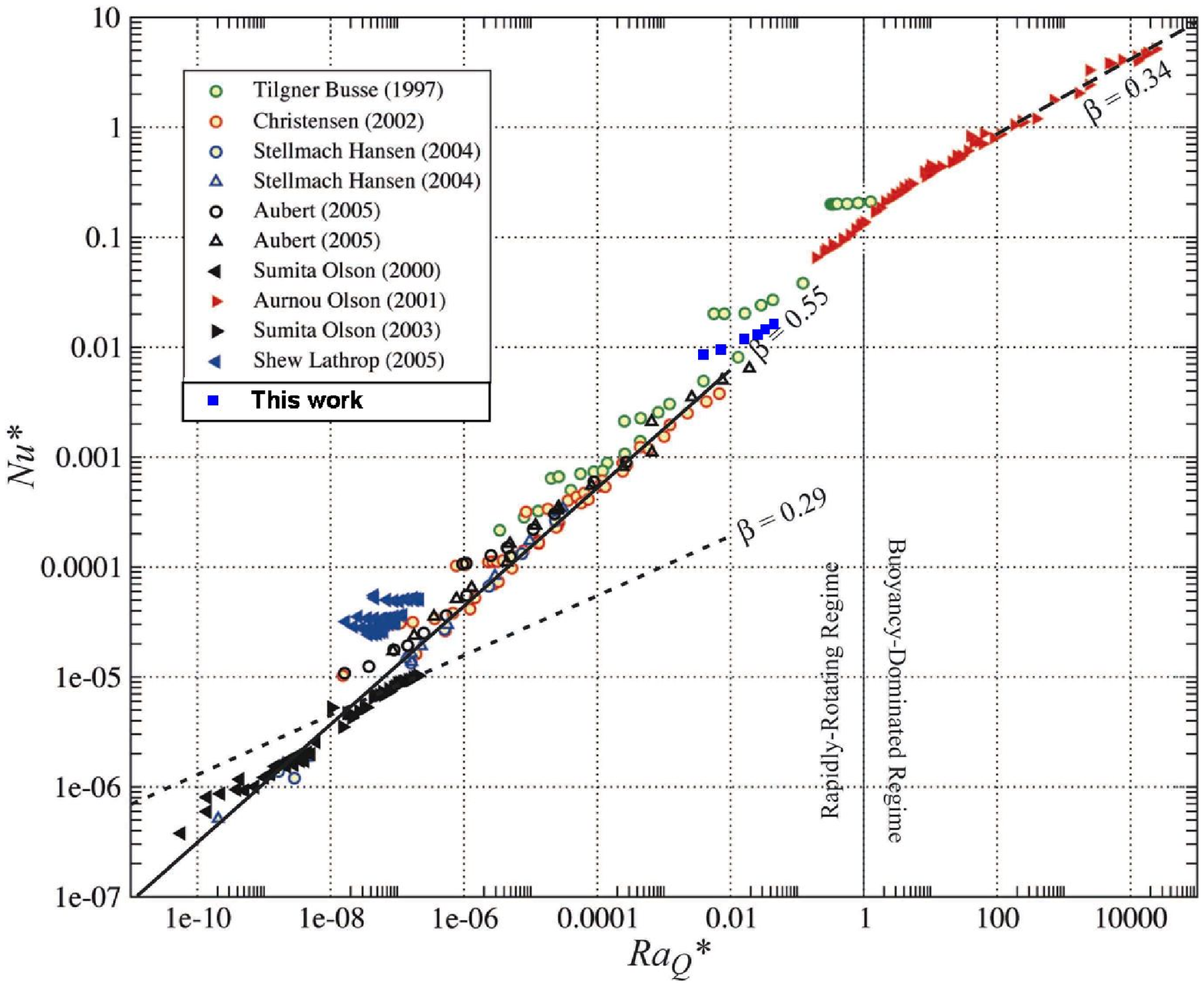}
  \end{center}
\end{figure}

\begin{figure}
  \begin{center}
    \epsfysize=9.0cm
    \leavevmode
    \epsfbox{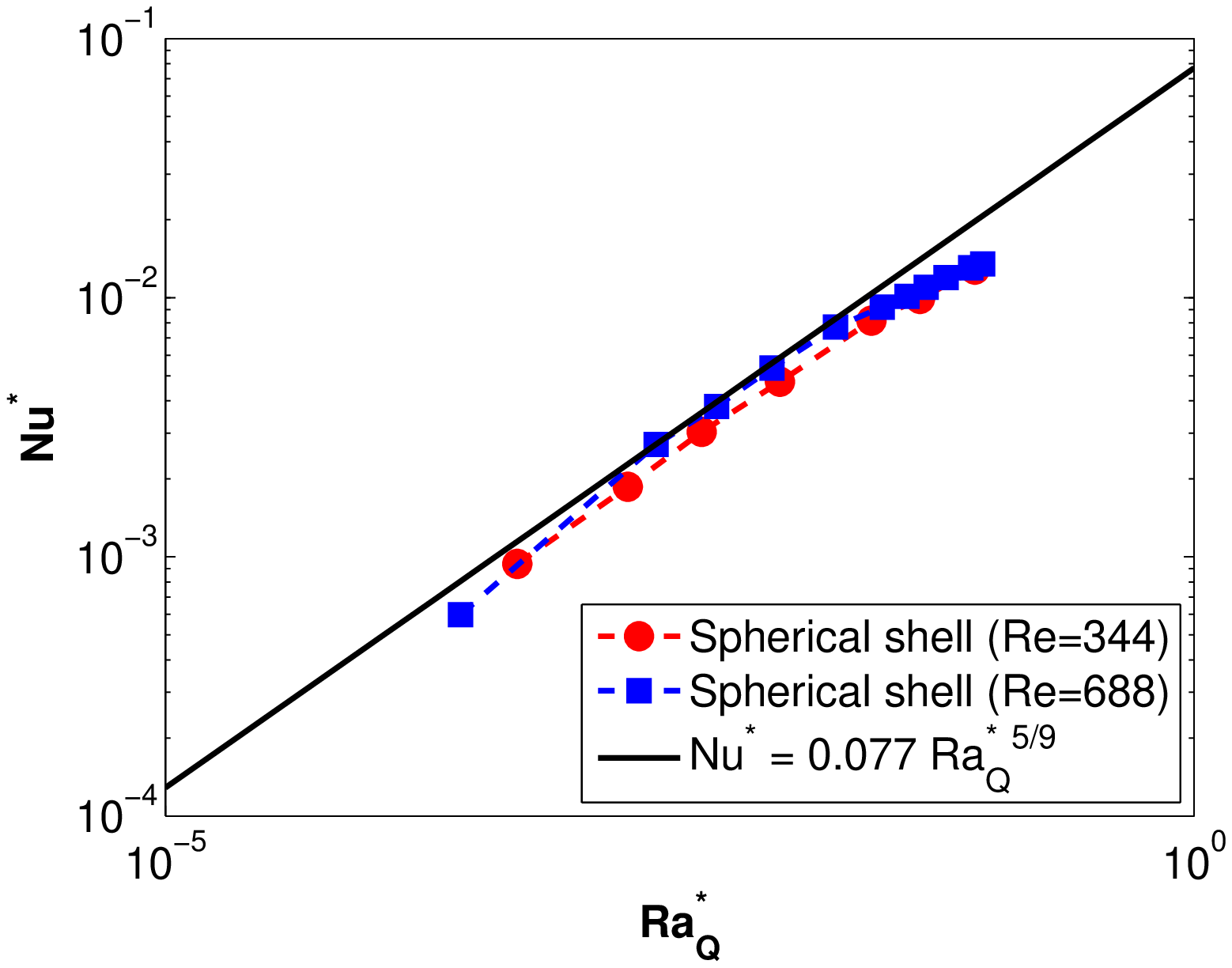}
    \caption{\it{(a) Upper figure. (b) Lower figure.}}
    \label{cebronfig2}
  \end{center}
\end{figure}

\begin{figure}
  \begin{center}
    \epsfysize=9.0cm
    \leavevmode
    \epsfbox{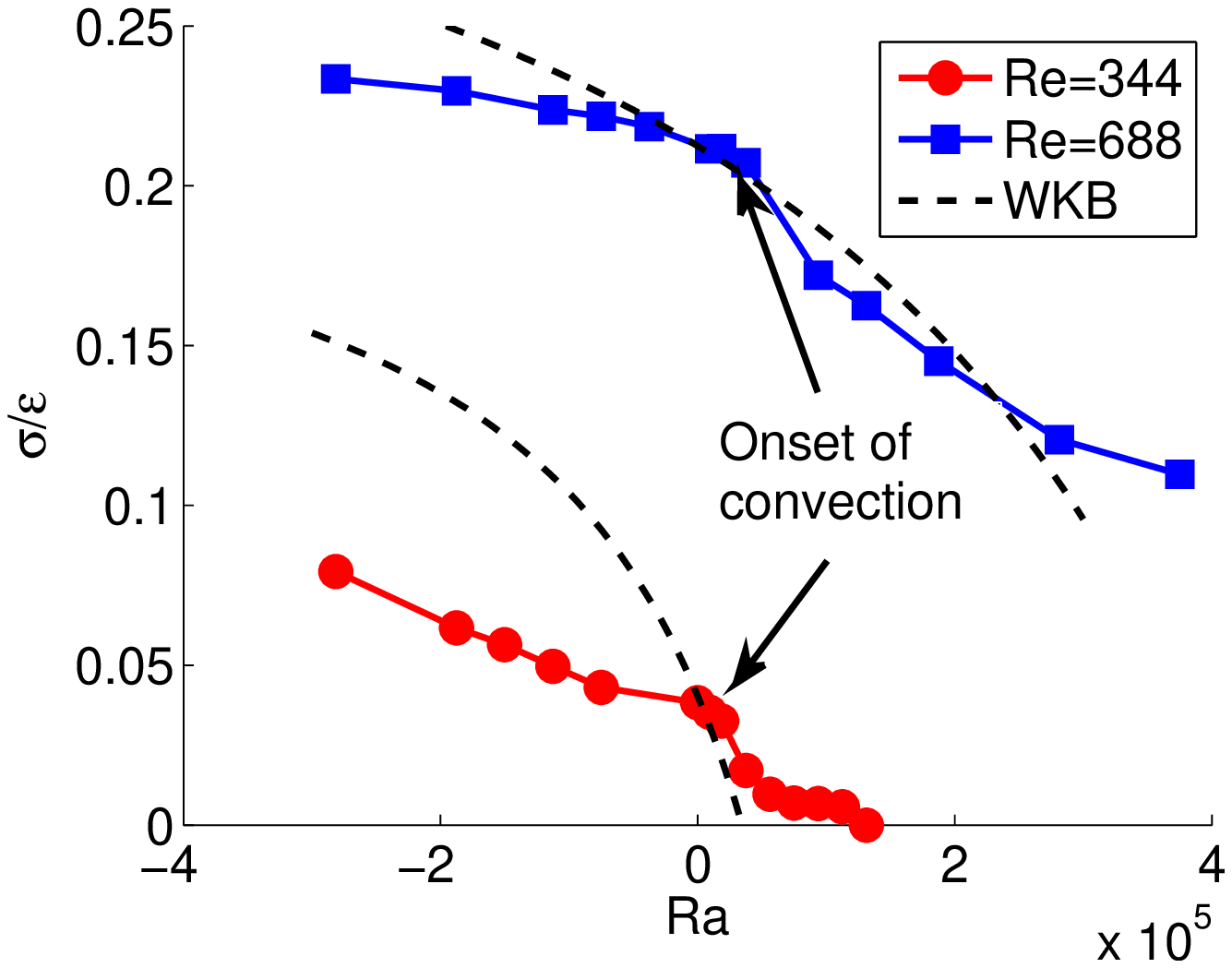}
  \end{center}
\end{figure}

\begin{figure}
  \begin{center}
    \epsfysize=9.0cm
    \leavevmode
    \epsfbox{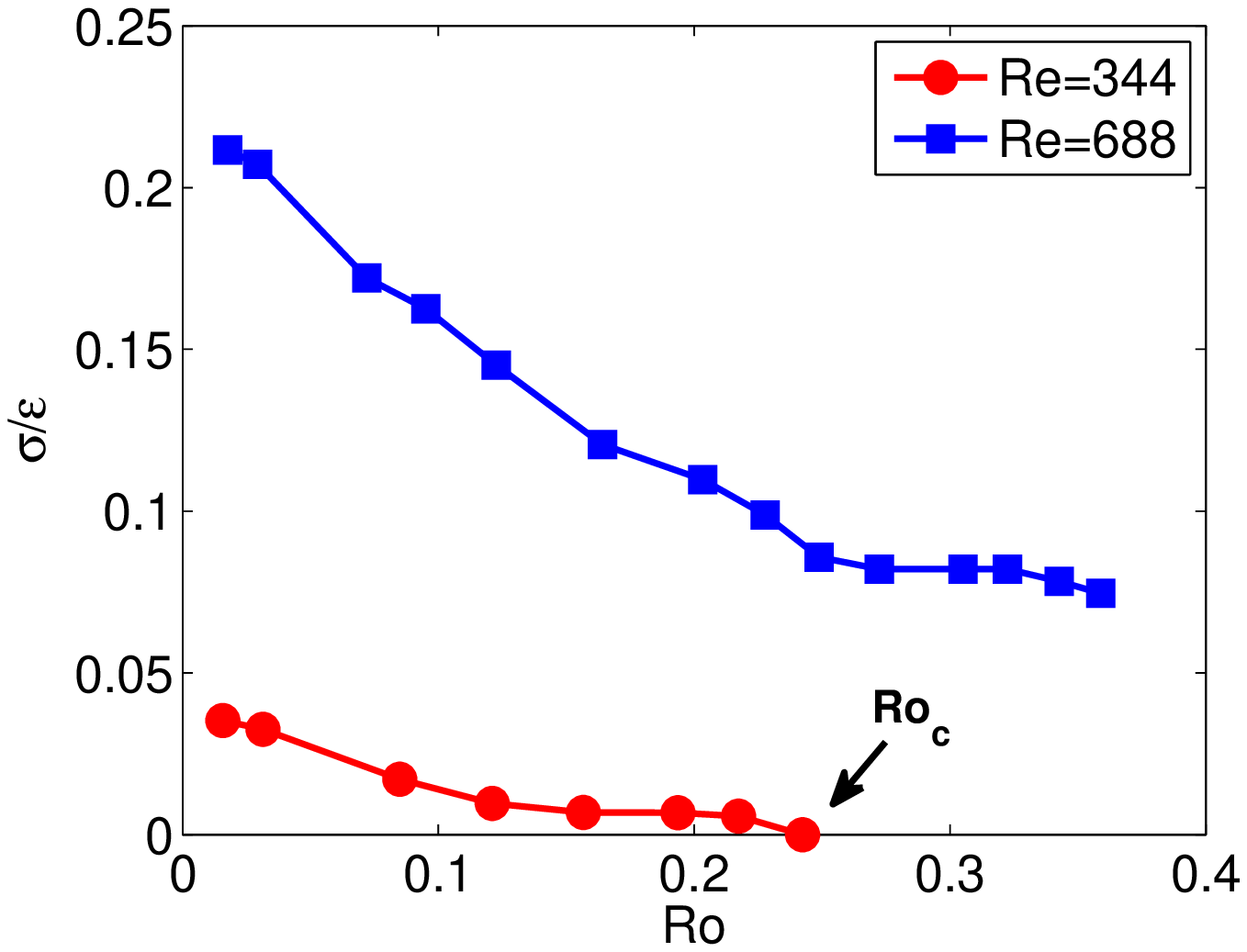}
    \caption{\it{(a) Upper figure. (b) Lower figure.}}
    \label{cebronfig4}
  \end{center}
\end{figure}

\begin{figure}
  \begin{center}
    \epsfysize=9.0cm
    \leavevmode
    \epsfbox{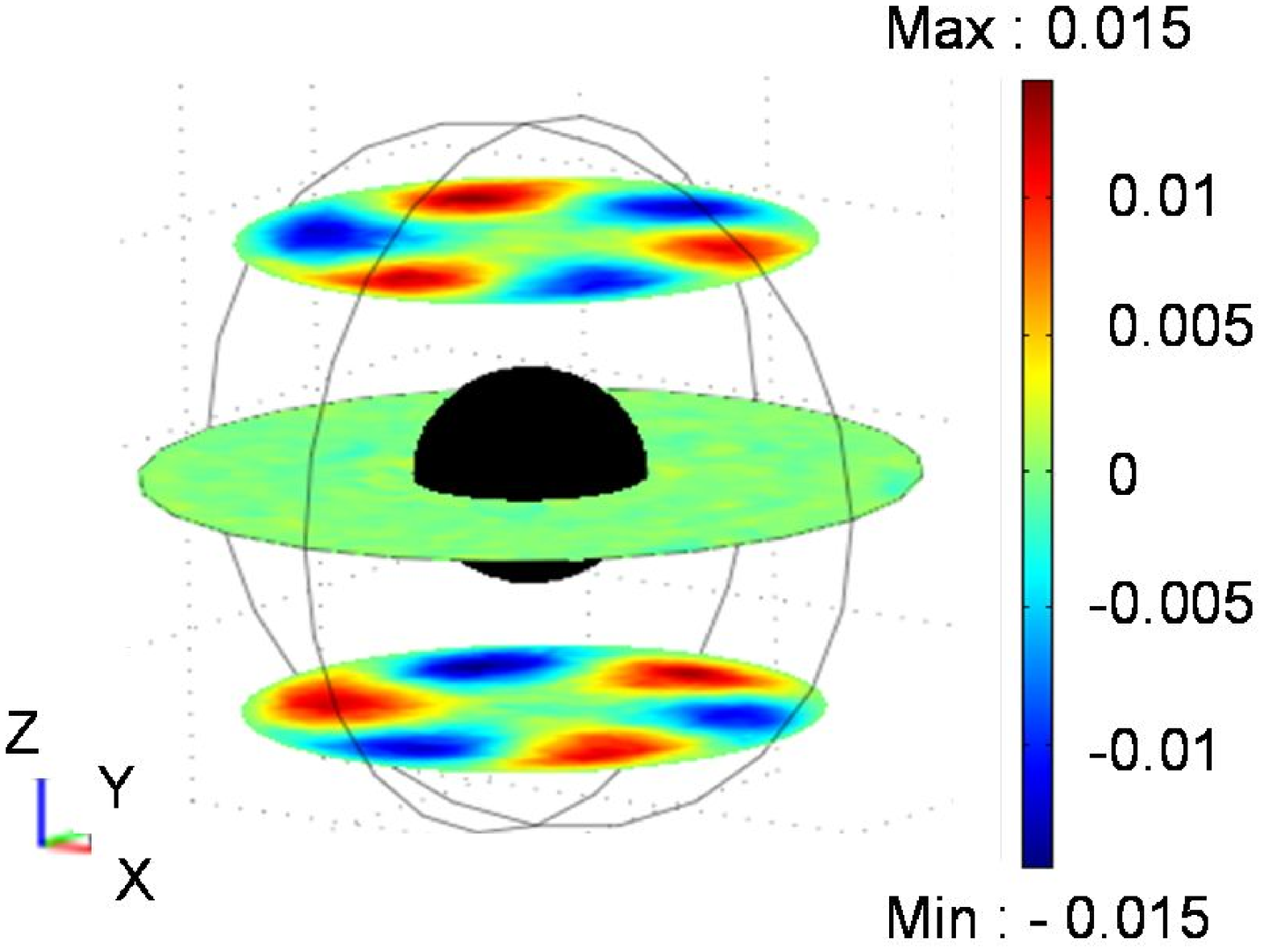}
  \end{center}
\end{figure}

\begin{figure}
  \begin{center}
    \epsfysize=9.0cm
    \leavevmode
    \epsfbox{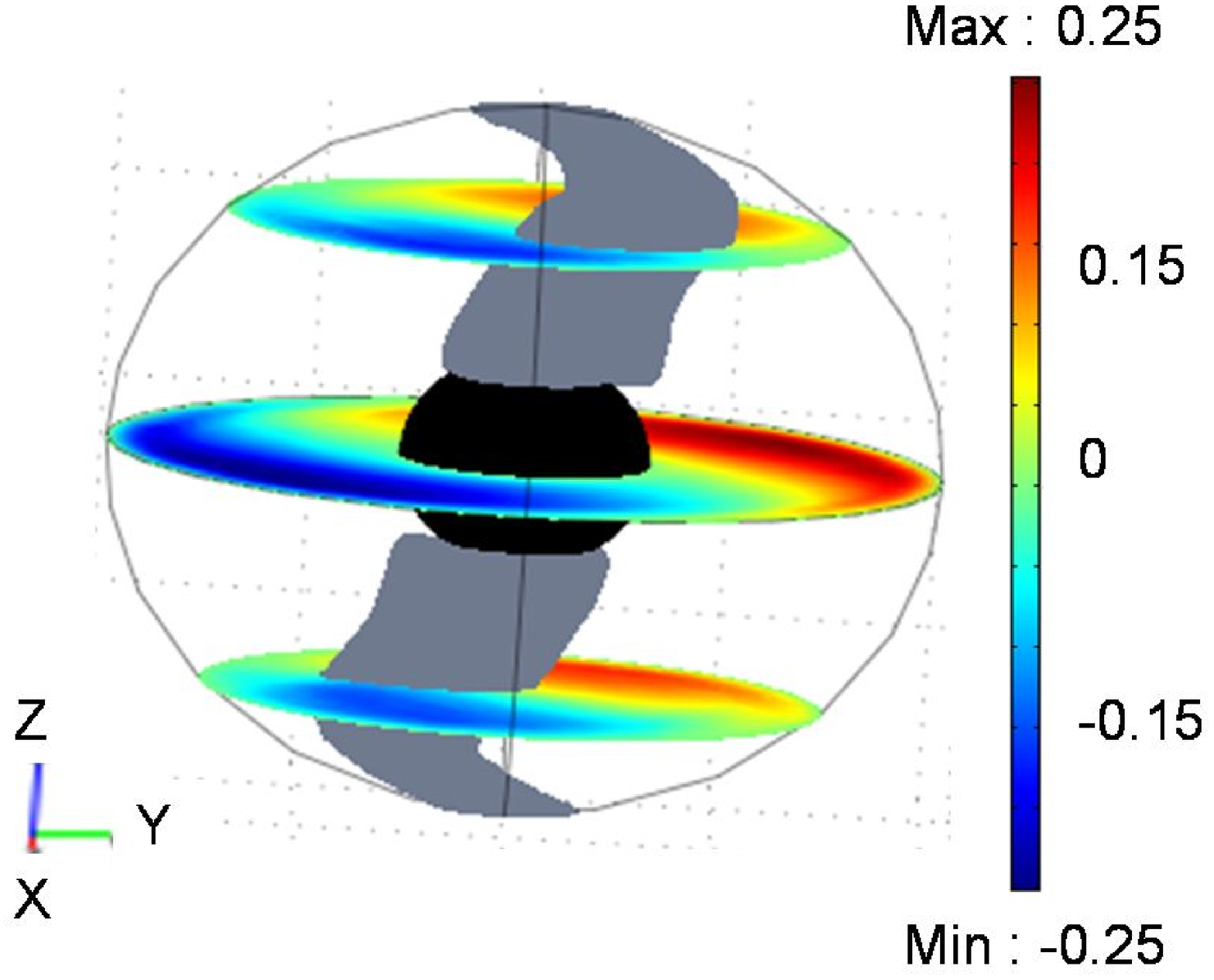}
    \caption{\it{(a) Upper figure. (b) Lower figure.}}
    \label{BusseIE}
  \end{center}
\end{figure}

\begin{figure}
  \begin{center}
    \epsfysize=9.0cm
    \leavevmode
    \epsfbox{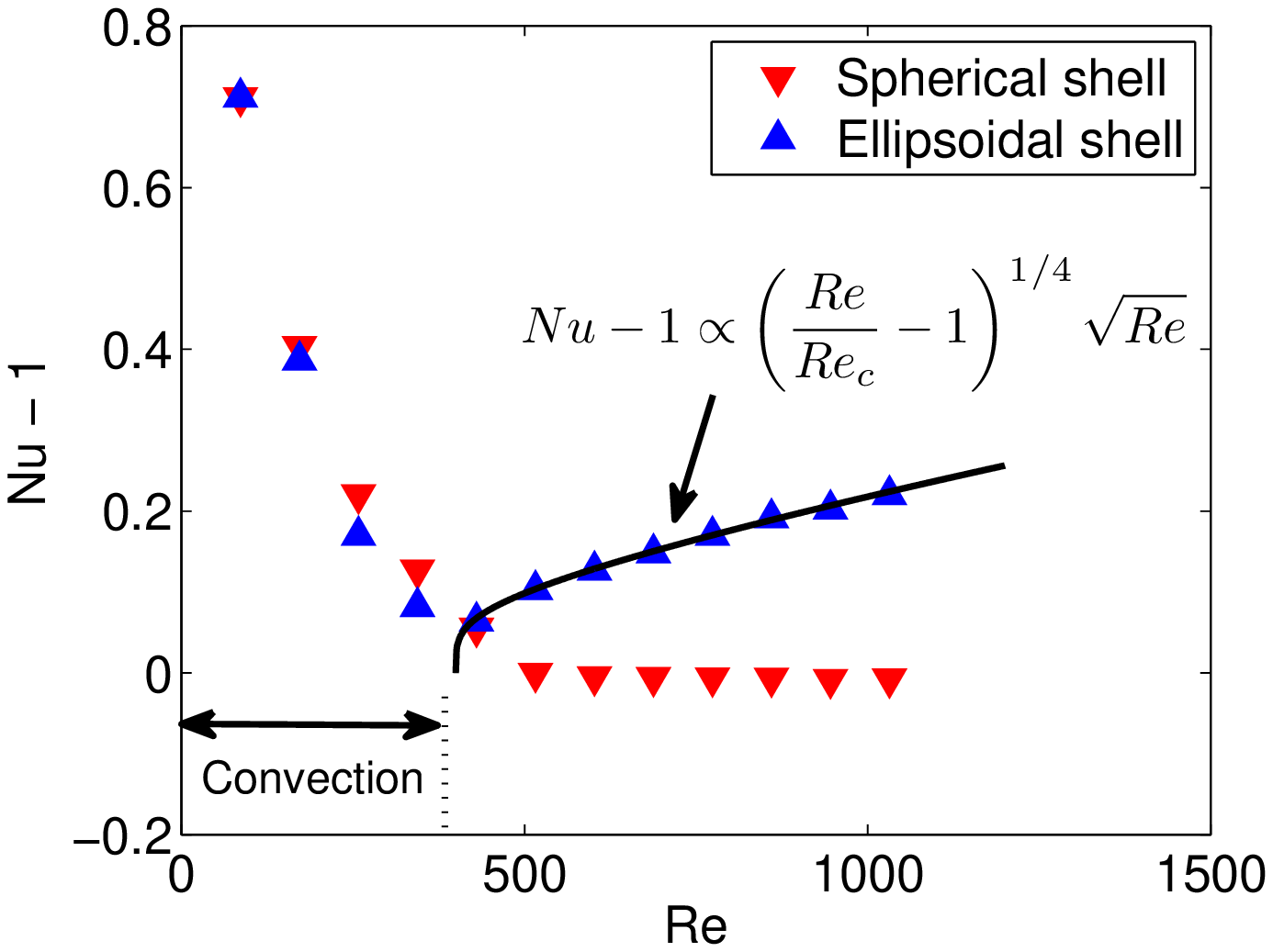}
  \end{center}
\end{figure}

\begin{figure}
  \begin{center}
    \epsfysize=9.0cm
    \leavevmode
    \epsfbox{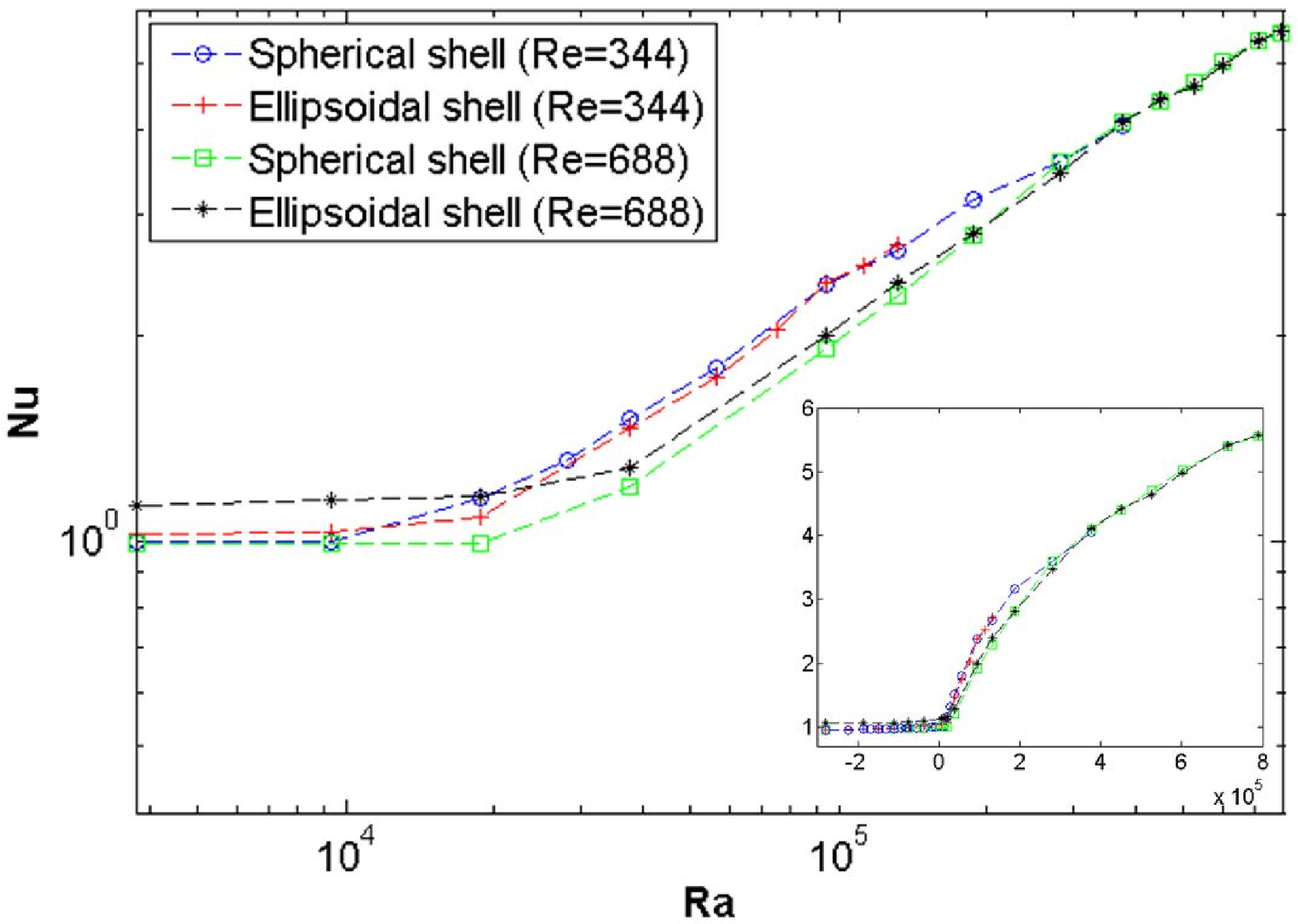}
    \caption{\it{(a) Upper figure. (b) Lower figure.}}
    \label{cebronfig5}
  \end{center}
\end{figure}

\begin{figure}
  \begin{center}
    \epsfysize=9.0cm
    \leavevmode
    \epsfbox{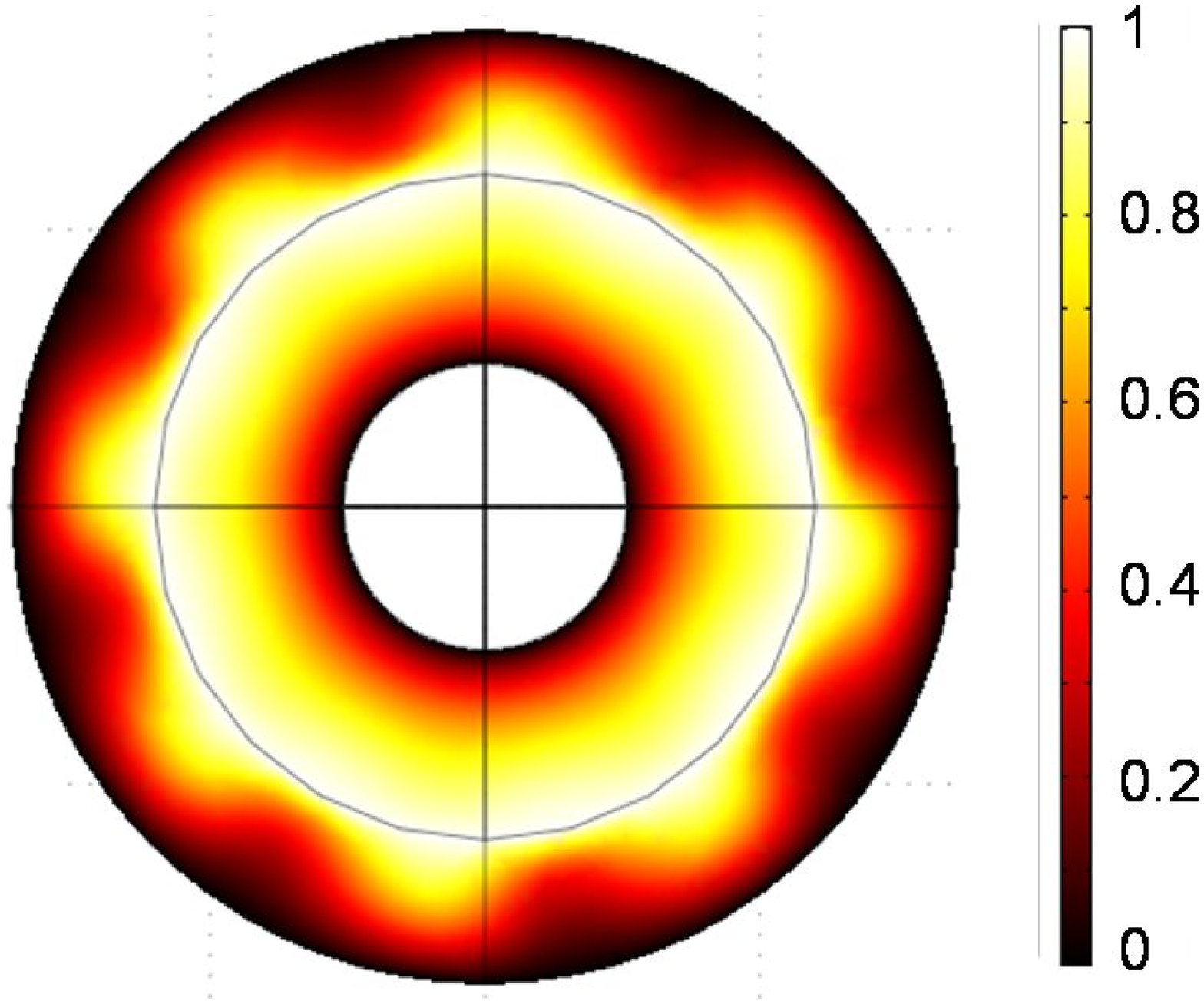}
  \end{center}
\end{figure}

\begin{figure}
  \begin{center}
    \epsfysize=9.0cm
    \leavevmode
    \epsfbox{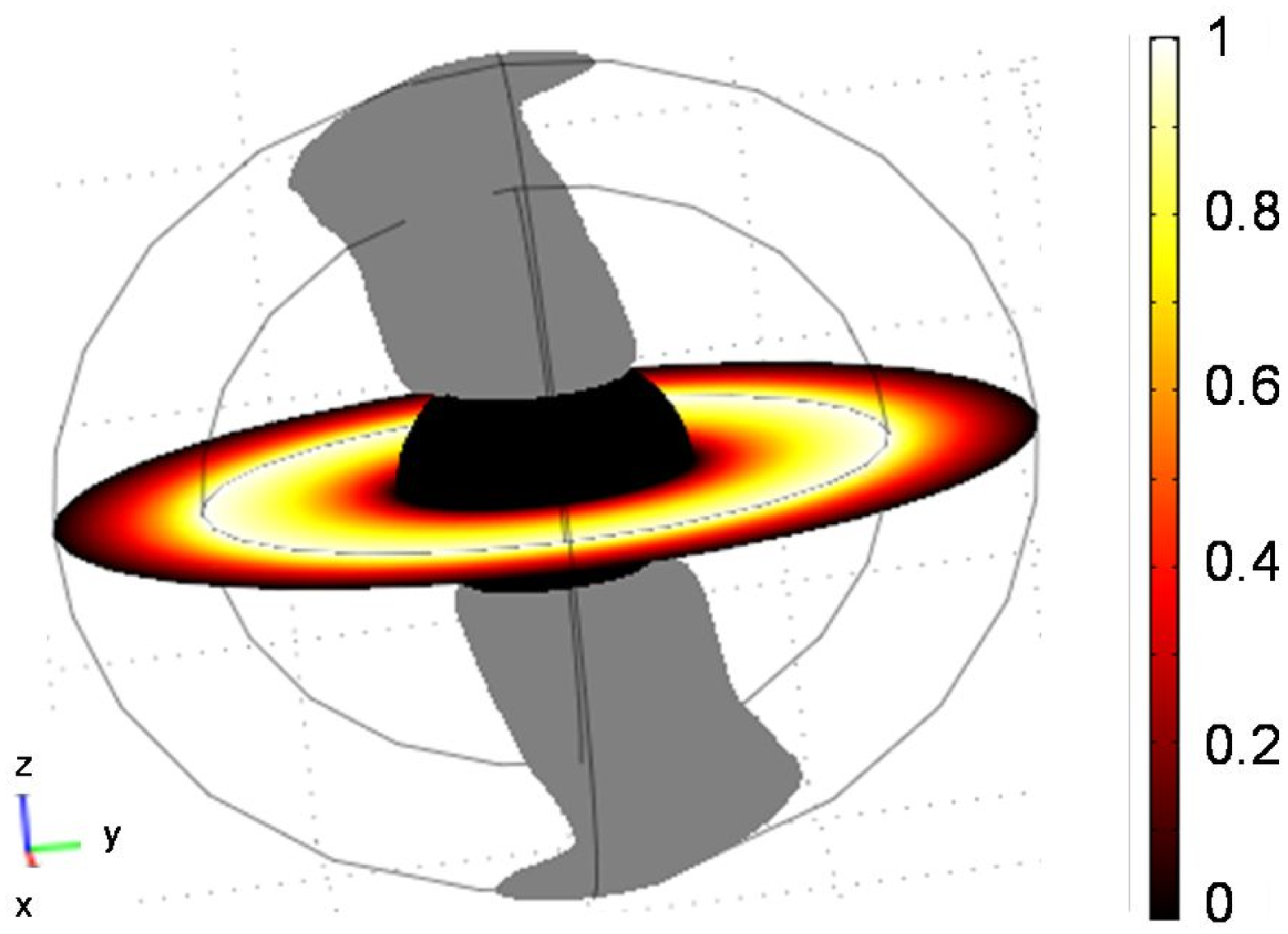}
    \caption{\it{(a) Upper figure. (b) Lower figure.}}
    \label{cebronfig7}
  \end{center}
\end{figure}

\end{document}